\documentclass[aps,prd,10pt,nofootinbib,twocolumn,superscriptaddress,showpacs]{revtex4-1} %
\usepackage[english]{babel} 
\usepackage[utf8]{inputenc}
\usepackage{latexsym} 
\usepackage{amsmath} 
\usepackage{amsfonts} 
\usepackage{amssymb} 
\usepackage{textcomp} 
\usepackage{graphicx}  
\usepackage{enumerate} 
\usepackage{array} 
\usepackage{mathrsfs}
\usepackage{amsthm}
\usepackage{bm,bbm}
\usepackage{bbold}  
\usepackage{multirow}
\usepackage{fancyref}
\usepackage{hyperref}
\usepackage{color}
\newcommand{\Eq}[1]{(\ref{#1})}

\newcommand{\be}{\begin{equation}}
\newcommand{\ee}{\end{equation}}
\newcommand{\ba}{\begin{eqnarray}}
\newcommand{\ea}{\end{eqnarray}}
\newcommand{\bs}{\begin{subequations}}
\newcommand{\es}{\end{subequations}}
\def\com{\color{magenta}}
\def\cob{\color{blue}}
\newcommand{\rmd}{d}
\newcommand{\rmi}{i}

\renewcommand{\dh}{d_\textsc{h}}

\newcommand{\arX}[1]{\href{http://arxiv.org/abs/#1}{{\ttfamily\com arXiv:#1}}}

\newcommand{\procsin}[5]{in \emph{#1}, ed.\ by #2 (#3, #4, #5)} % for Singapore
\newcommand{\doin}[6]{\href{http://dx.doi.org/#1}{\cob #2\ #3 {\bf #4}, #5 (#6)}}

\newcommand{\doinn}[5]{\href{http://dx.doi.org/#1}{{\cob #2 {\bf #3}, #4 (#5)}}}
\newcommand{\doij}[5]{\href{http://dx.doi.org/#1}{{\cob #2 {#3} (#5) #4}}}

\newcommand{\tia}[1]{}

\def\ra{\rightarrow}

\def\ove{\overline}

\def\cL{\mathcal{L}}
\def\p{\partial}
\def\a{\alpha}

\def\s{\sigma}

\def\l{\lambda}

\def\cP{\mathcal{P}}
\def\om{\omega}

\def\ds{d_{\rm S}}
\newcommand{\Pl}{{\text{\tiny Pl}}}

\begin{document}

\title{Particle-physics constraints on multifractal spacetimes}
\author{Gianluca Calcagni}
\affiliation{Instituto de Estructura de la Materia, IEM-CSIC, Serrano 121, 28006 Madrid, Spain}
%\email{calcagni@iem.cfmac.csic.es}
\author{Giuseppe Nardelli}
\affiliation{Dipartamento di Matematica e Fisica, Università Cattolica del Sacro Cuore, via Musei 41, 25121 Brescia, Italy}
\affiliation{TIFPA -- INFN, c/o  Dipartimento di Fisica, Universit\`a di Trento, 38123 Povo (TN), Italy}
%\email{giuseppe.nardelli@unicatt.it}
\author{David Rodr\'{\i}guez-Fern\'andez}
\affiliation{Departamento de F\'isica, Universidad de Oviedo, Avda.~Calvo Sotelo 18, 33007, Oviedo, Spain}
\affiliation{Departamento de F\'{\i}sica Te\'orica II, Universidad Complutense de Madrid, Parque de las Ciencias 1, 28040 Madrid, Spain}
%\email{rodriguezferdavid@uniovi.es}

\begin{abstract}
We study electroweak interactions in the multiscale theory with $q$-derivatives, a framework where spacetime has the typical features of a multifractal. In the simplest case with only one characteristic time, length, and energy scale $t_*$, $\ell_*$, and $E_*$, we consider (i) the muon decay rate and (ii) the Lamb shift in the hydrogen atom, and constrain the corrections to the ordinary results. We obtain the independent absolute upper bounds (i) $t_* < 10^{-13}{\rm s}$ and (ii) $E_*>35\,\text{MeV}$. Under some mild theoretical assumptions, the Lamb shift alone yields the even tighter ranges $t_*<10^{-27}\,{\rm s}$, $\ell_*<10^{-19}\,{\rm m}$, and $E_*>450\,\text{GeV}$. To date, these are the first robust constraints on the scales at which the multifractal features of the geometry can become important in a physical process.
\end{abstract}

\date{November 3, 2015}


% 11.10.Kk Field theories in dimensions other than four
% 12.60.-i	Models beyond the standard model
% 05.45.Df Fractals
% 04.60.-m Quantum gravity

\preprint{\doin{10.1103/PhysRevD.93.025005}{PHYSICAL REVIEW}{D}{93}{025005}{2016} \hspace{9cm} \arX{1512.02621}}

\maketitle

%%%%%%%%%%%%%%%%%%%%%%%%%%%%%%%%%%%%%%%%%%%%%%%%%%%%%%%%%%%%%%%%%%%%%%%
%%%%%%%%%%%%%%%%%%%%%%%%%%%%%%%%%%%%%%%%%%%%%%%%%%%%%%%%%%%%%%%%%%%%%%%

The variety of theories of quantum gravity proposed in the last 30 years has highlighted two facts. First, that there are many languages and mathematical tools with which one can describe consistent quantum geometries and that, while some of these approaches are mutually exclusive, others can be related in nontrivial ways or even embedded into one another. Second, that despite their differences these approaches (including causal dynamical triangulations in the ``de Sitter'' phase, asymptotic safety, Hor\v{a}va--Lifshitz gravity, nonlocal gravity, loop quantum gravity and spin-foams, noncommutative spacetimes, quantum black holes, and more \cite{qgrefs}) share some distinct features. One of the most striking phenomena one comes across the landscape of quantum-gravity models is \emph{dimensional flow}, the change of the dimension of spacetime (whenever a notion of spacetime emerges meaningfully in each approach) with the probed scale \cite{tH93Car09,fra1}. In any approach to quantum gravity, the effective dimension flows to four at low energies and large scales, where general relativity is an impeccable description of geometry. At small scales, however, the spectral dimension $\ds$ of spacetime can attain a completely different value, usually equal to or smaller than 2. The transition between the two regimes varies depending on the model but it is usually difficult to have it under full analytic control. For this and other reasons, most of the physical consequences of dimensional flow in contexts as diverse as quantum field theory (QFT) and cosmology remain elusive.

Nevertheless, it is surprisingly easy to reproduce \emph{ad hoc} the dimensional flow found in various quantum-gravity theories \cite{frc64}. This is achieved by placing a field theory $\cL[\phi_{\mu\nu\dots}]$ on a geometry with action $S=\int\rmd^4x\,v\,\cL$, where $v(x)=v_*(t)\,v_*({\bf x})$ is a \emph{non}dynamical profile (unrelated to the volume density $\sqrt{-g}$ in curved spacetimes) with
\be\label{binnosc}
v_*(t)=\left(1+\left|\frac{t}{t_*}\right|^{\a-1}\right)\!,\quad v_*({\bf x})=\prod^{3}_{i=1}\left(1+\left|\frac{x^i}{\ell_*}\right|^{\a-1}\right)\!.
\ee
The parameter $0<\a<1$ is called fractional exponent (it can be different along different directions, but here this complication is not necessary). This geometry has Hausdorff dimension $\dh=4$ at large scales $\Delta\ell\gg \ell_*$ and late times $\Delta t\gg t_*$, while $\dh=4\a<4$ at small scales and early times. The spectral dimension $\ds$ has a similar behavior and, in particular, $\ds=2$ for $\a=1/2$. Perhaps, the most remarkable property of Eq.~\Eq{binnosc} is that it is not just a heuristic profile useful to fit numerical data points or asymptotic regimes; it also represents \emph{uniquely} the continuum approximation of random multifractals \cite{NLM,frc2}. 

Based on this observation, the theory of \emph{multiscale spacetimes} has been proposed recently \cite{fra1,frc1,frc2} (see \cite{AIP,frc11} for reviews). By embedding dimensional flow by default, it provides the means for an agile study of multifractal properties of anomalous geometries, even not related to quantum gravity. Early attempts to construct field theories on specific fractals \cite{fracs} were not manageable beyond a first formal stage of development. The versatility of multiscale models has permitted us to extract abundant phenomenology and to satisfy our curiosity about how the world would be if the fabric of spacetime was fractal. Regarded as stand-alone proposals, the four extant multiscale theories (with normal, weighted, fractional, and $q$-derivatives) should address a number of questions, including the existence and magnitude of exotic effects across all scales and the possibility to constrain the theory with present or forthcoming experiments. Much progress has been made regarding theoretical aspects such as accelerating cosmological solutions \cite{frc11} and the renormalizability of QFTs \cite{frc9}, but there is still little contact with observations. 

In this paper, we examine the multiscale model with so-called $q$-derivatives, which is more intuitive and under better control than the others. After introducing the $SU(2)\otimes U(1)$ QFT of electroweak interactions, we consider the decay rate of the muon and the Lamb-shift effect in the hydrogen atom, and we ascribe the experimental uncertainty of the most recent measurements to multifractal effects. This strategy, originally adopted in early toy models of dimensional regularization \cite{ZSMuSScM}, is crude but quite effective, since it will allow us to place {the first absolute bounds on the time, length, and energy scales characterizing the geometry. The full Standard Model with the inclusion of strong interactions, the details of the calculations, and the (much more involved) study of the multiscale theory with weighted derivatives will be presented in a companion paper \cite{frc13}. In the following, $c=1=\hbar$ and we ignore curvature effects.

\emph{The theory.}---In the multiscale theory with $q$-derivatives, the volume element $\rmd^4x$ in any action $S$ is replaced by a measure $\rmd^4x\,v(x)$ which depends on some characteristic time and length scales $t_*$ and $\ell_*$. The length $\ell_*$ determines the difference between ``infrared'' and ``ultraviolet,'' while $t_*$ sets the time scale below which a physical process (e.g., a particle scattering) feels the imprint of fractal geometry. It can also be interpreted as the end of an early cosmological era dominated by fractal effects \cite{frc8}. The simplest possible multiscale measure $v(x)$ is the binomial profile \Eq{binnosc} and corresponds to a \emph{random} multifractal. On the other hand, many \emph{deterministic} multifractals \cite{NLM,frc2} are approximated by the measure $v_{\rm log}(x) = [1+|{t}/{t_*}|^{\a-1}F_\om(\ln|t/t_\infty|)]\prod_i[1+|{x^i}/{\ell_*}|^{\a-1}F_\om(\ln|x^i/\ell_\infty|)]$. An analysis of the spectral and walk dimension shows that these spacetimes have all the main properties of multifractals \cite{frc7}. The modulation factor $F_\om(\ln|x/\ell_\infty|)=A\,\cos(\om\ln|x/\ell_\infty|)+B\,\sin(\om\ln|x/\ell_\infty|)$ features logarithmic oscillations with frequency $\om$, representing a discrete scale invariance $x\to \exp(2\pi n/\om)\,x$ of the geometry. The fundamental scales $t_\infty$ and $\ell_\infty$ are much smaller than $t_*$ and $\ell_*$. In \cite{ACOS}, it was argued that if one identifies these quantities with the Planck scales $t_\infty=t_\Pl$ and $\ell_\infty=\ell_\Pl$, then multiscale spacetimes with the measure $v_{\rm log}(x)$ provide the natural completion of $\kappa$-Minkowski spacetime. Theoretical aspects of the measure, such as changes in presentation [$v(x)\to v(x-\bar x)$], are further discussed in \cite{frc11,frc13}. Log oscillations have no \emph{direct} impact on the physics at LHC scales, but we will reintroduce them at the end of the paper for a crucial unit conversion.

For any measure $v(x)$, relativistic actions with $q$-derivatives are invariant under the nonlinear $q$-Poincaré symmetries ${q'}^\mu(x^\mu)=\Lambda_\nu^{\ \mu}q^\nu(x^\nu)+a^\mu$, where $q^\mu(x^\mu):=\int^{x^\mu}\rmd {x'}^\mu\,v_\mu ({x'}^\mu)$ are called \emph{geometric coordinates}. The measure can be rewritten as $\rmd^4x\,v(x)=\rmd^4 q(x)=\rmd q^0(x^0)\dots \rmd q^{3}(x^3)$. For the binomial measure \Eq{binnosc}, the geometric coordinate in the time direction is
\be\label{binomialm2}
q_*(t) = t+t_*\frac{{\rm sgn}(t)}{\a}\left|\frac{t}{t_*}\right|^{\a}.
\ee 
The expression of the measure $\rmd^4p(k)$ in momentum space and of its coordinates $p^\mu(k^\mu)$ is universal and valid for arbitrary geometric coordinates \cite{frc11}: $p^\mu(k^\mu)=1/q^\mu(1/k^\mu)$, where all the time-length scales appearing in $q^\mu$ are replaced by a hierarchy of energy-momentum scales $\{E_*, {\bf k}_*\}$. Here we will be interested in the energy geometric coordinate associated with Eq.\ \Eq{binomialm2}:
\be
p_*(E) = \left[\frac{1}{E}+\frac{{\rm sgn}(E)}{E_*\a}\left|\frac{E_*}{E}\right|^{\a}\right]^{-1}.\label{bino2}
\ee
The dynamics of any system of interest (Einstein gravity, the Standard Model, and so on) is easily defined. It is the usual one under the replacement
\be\label{xq}
x\to q(x)\,,\qquad k\to p(k)
\ee
everywhere. A Lagrangian $\cL[\p_x,\phi_{\mu\nu\dots}(x)]$ with generic fields $\phi_{\mu\nu\dots}$ becomes $\cL\{\p_{q(x)},\phi_{\mu\nu\dots}[q(x)]\}$. Multiscale spacetimes are a framework where $q$-measurements are performed with instruments which adapt with the observation scale \cite{fra7}. This adaptation is encoded in the structure of the integration measure and of differential operators, where characteristic time, length, and energy scales appear. Measurement units for the coordinates must be specified. Time and spatial coordinates scale as lengths, $[t]=-1=[x^i]$, which set our clocks and rods. On the other hand, geometric coordinates have an anomalous scaling with respect to these clocks and rods: in the ultraviolet, one has $q\propto x^\a$ and an anomalous scaling for $\a\neq 1$. By definition of the theory, time intervals, lengths, and energies are physically measured in the frame with coordinates $x^\mu$ ($k^\mu$ in momentum space), where coordinate transformations are described by the nonlinear $q$-Poincaré symmetries. Equation \Eq{xq} (which is not a coordinate transformation) governs the passage between a frame $\{x^\mu, k^\mu\}$, called \emph{fractional picture}, where physical observables should be computed and a frame $\{q^\mu, p^\mu\}$, called \emph{integer picture}, where intermediate steps of such calculations can be carried on. An example of the nontriviality of the $q$-theory, due to the exotic structure \Eq{bino2} of momentum space, is the primordial cosmological spectrum of inflation \cite{frc11}.

Since the frame where physical measurements are performed is established uniquely, it is possible to predict deviations of particle-physics observables from the standard lore. However, when the action is written explicitly in $x$ coordinates, it resembles an inhomogeneous field theory in ordinary spacetime with noncanonical kinetic terms and nonconstant couplings. For example, the action of a static real scalar field with polynomial potential in one dimension would be $S_\phi=-\int\rmd q\,[(\p_q\phi)^2/2+\sum_n\l_n\phi^n]=-\int\rmd x\{(\p_x\phi)^2/[2v(x)]+\sum_n\bar\l_n(x)\,\phi^n\}$, where $\bar\l_n(x)=v(x)\l_n$. Since we do not know how to define a fully predictive perturbative QFT with varying couplings \cite{frc9} or inhomogeneous kinetic terms, it is necessary to perform all calculations in geometric coordinates. At the end of the calculation, one must return to the physical frame to interpret the results correctly. Any ``time'' or ``spatial'' interval or ``energy'' predicted in the integer picture are not a physical time or spatial interval or energy, since they are measured with $q$-clocks, $q$-rods or $q$-detectors. The results must be reconverted to physical measurement units in the fractional picture.

Therefore, in the case of the $SU(3)\otimes SU(2)\otimes U(1)$ Standard Model (discussed in detail in \cite{frc13}), it is sufficient to go to the integer picture, where it coincides with the usual Standard Model that can be found in textbooks, and respectfully borrow any theoretical result we wish to compare with experiments. The only nontrivial step is the unit reconversion. We give two examples of this procedure: the muon lifetime and the Lamb shift.

\emph{Muon decay rate.}---In ordinary flat spacetime, the probability distribution for the energy $E$ of an unstable particle with mass $m$ is governed by the Breit--Wigner  distribution $f_{\rm BW}(E)\propto\Gamma/[(m^2-E^2)^2+ (m\Gamma)^2]$, where $\Gamma$ is called decay width. $f_{\rm BW}(E)$ is the square of the quantum amplitude describing the decay of the resonance, which, in turn, is proportional to the propagator of the particle. The decay width can be calculated explicitly for the unstable particles appearing in the Standard Model. To a scattering process described by a one-particle initial state $|{\rm i}\rangle$ and a many-particle final state $|{\rm f}\rangle$, one computes the Feynman amplitude $\langle {\rm f}\vert {\rm i}\rangle$ up to a certain perturbative order. From the transition probability $\cP({\rm i}\ra {\rm f}) \propto |\langle {\rm f}\vert {\rm i}\rangle|^2=f_{\rm BW}$, the decay rate $\Gamma$ for the resonance $|{\rm i}\rangle$ is then extracted. In the case of the muon, the process is $\mu^- \ra e^- \ove{\nu}_e \nu_{\mu}$ and it is mediated by a gauge boson $W$. Neglecting the masses of the electron $e^-$ and the neutrino $\nu_e$, one has
\be \label{gamma0}
\Gamma = \frac{G^2_{\rm F} m_{\rm mu}^5}{192\pi^3}+\dots\,,
\ee
where $G_{\rm F}$ is Fermi's constant, $m_{\rm mu}$ is the muon mass and the ellipsis denotes loop corrections to the tree-level contribution. The mean lifetime $\tau_{\rm mu}$ of the resonance is identified with the inverse of $\Gamma$, $\tau_{\rm mu}=\tau_0:=1/\Gamma$.

In the theory with $q$-derivatives, one works in the integer picture and obtains \Eq{gamma0}. However, $\Gamma$ is a composite object no longer equal to the inverse of the muon lifetime. From the form of the propagator $\propto [p^2(k)+m_{\rm mu}^2+\rmi m_{\rm mu}\Gamma]^{-1}$, it is natural to make the identification $\Gamma=p_*(1/\tau_{\rm mu})=1/q_*(\tau_{\rm mu})$, and the physically observed muon lifetime is found by inverting the relation
\be\label{ansq1}
q_*(\tau_{\rm mu}) = \tau_{\rm mu}+\frac{t_*}{\a}\left(\frac{\tau_{\rm mu}}{t_*}\right)^{\a}=\tau_0=\frac{1}{\Gamma}\,.
\ee
The replacement of $\tau_{\rm mu}=\tau_0$  with this formula gives a characteristic prediction that can be compared with that in standard Minkowski spacetime. The muon lifetime is not observed directly. Experiments determine the Fermi constant $G_{\rm F}= 1.1663787(6)\times 10^{-5}\,\text{GeV}^{-2}$ and the muon mass $m_{\rm mu} = 105.6583715(35) \,\text{MeV}$ \cite{pdg}, where the numbers in round brackets denote the first nonzero digits of the $1\s$-level experimental error and apply to the last figure(s) given in the number. Using Eq.~\Eq{gamma0}, one has $\tau_0=2.1969811(22)\times 10^{-6}\,{\rm s}$ for $\mu^-$ \cite{pdg}. The lifetime of $\mu^+$ is almost the same and we can ignore the difference. If we knew both $\a$ and $t_*$, we would invert Eq.~\Eq{ansq1} and find the multiscale prediction for $\tau_{\rm mu}$. As we do not, we opt for a different approach. We assume realistically that $t_*$ is small enough so that the scale-dependent part of the measure is small and $\tau_{\rm mu}\approx \tau_0$ to a very good approximation. Then, we account for all the experimental error $\delta\tau\approx 6.6\times 10^{-12}\,{\rm s}$ at the $3\s$-level as setting an upper limit on the effects of anomalous geometry: $(t_*/\a)({\tau_0}/{t_*})^{\a}<\delta\tau$, implying that
\be\label{tstfin}
t_*< \left(\frac{\a\delta\tau}{\tau_0^{\a}}\right)^{\frac{1}{1-\a}}\,.
\ee
Computing \Eq{tstfin} as a function of $0<\a<1$, we find that the maximum $t_*$ is attained for $\a\approx 0.06$. This value of $\a$ has no special meaning in the theory but it sets the \emph{absolute} upper bound $t_* < t_{\rm max}=10^{-13}\,{\rm s}$, independent from any other parameter of the model. To get stronger constraints, one can pick
the central value $\a=1/2$, which is special not only for its position in the middle of the prescribed interval $(0,1)$ \cite{frc1}, but also because it gives 2 dimensions in the ultraviolet \cite{frc7}, a feature much celebrated in quantum gravity \cite{tH93Car09,fra1} (other theoretical justifications for $\a=1/2$ can be found in \cite{frc2,frc1}). In this case, the allowed range $t_*<t_{\rm max}^{(\a=1/2)}$ is lowered by 5 orders of magnitude, $t_{\rm max}^{(\a=1/2)}= 5\times 10^{-18}\,{\rm s}$.

\emph{Lamb shift.}---According to quantum electrodynamics, the spectrum of the electron in the hydrogen atom depends on the principal and orbital-momentum quantum numbers. The emission and absorption of virtual photons by electrons and the production of virtual electrons in internal photon lines in Feynman diagrams give rise to a splitting of the spectral lines of different spin orbitals and, in particular, to a shift in the energy of the ${}^2P_{1/2}$ state with respect to the ${}^2S_{1/2}$ state. The measurement of this shift is one of the precision tests of perturbative QFT and has by now been verified for a number of light hydrogenic atoms \cite{EGSKar05}. For instance, the measured shift $\Delta E= E_{2S}-E_{2P}$ in hydrogen is \cite{Sch99}
\be
\Delta E      =4.37489(1)\times 10^{-6}\,\text{eV}\,.\label{hyd}
%\Delta E^{\text{D}}      &=& 1059.234(3)h\,\text{MHz}\,,\label{deu}\\
%\Delta E^{\text{He}\!^+} &=& 14041.1(2)h\,\text{MHz}\,,\label{hel}
\ee
The theoretical value predicted by quantum electrodynamics is in excellent agreement with observations.

In the theory with $q$-derivatives, the theoretical calculation of the radiative corrections to the Lamb shift is identical to the ordinary one upon the replacement $E\to p_*(E)$. Since we expect $E_*$ to be much larger than the characteristic energy scale involved in these experiments, we can make the approximation $E_*\gg E$ in \Eq{bino2}. A check \emph{a posteriori} will confirm this step. For $0<\a<1$, one has $p_*(E) \simeq E-(|E|/\a)|{E_*}/{E}|^{\a-1}$, so that the difference $\Delta p_*(E)=p_*(E_1)-p_*(E_2)$ between geometric energies is related to the difference $\Delta E=E_1-E_2$ between energies by $\Delta p_*(E) \simeq \Delta E+[(2-\a)/\a]|{E_1}/{E_*}|^{1-\a}(|E_2|-|E_1|)$, where we have used the fact that, for the levels $2S$ and $2P$ of hydrogenic atoms, $\Delta E/E_1\sim \Delta E/E_2\ll 1$. The expansion $x^a-1=a(x-1)+O[(x-1)^2]$ then applies. Identifying $E_1=E_{2S}$ and $E_2=E_{2P}$ with the energy of, respectively, the $2S_{1/2}$ and $2P_{1/2}$ state and noting that both $E_{2S}$ and $E_{2P}$ are negative, the relation between geometric and physical Lamb shift is
\be\label{pfin}
\Delta p_*(E) \simeq \Delta E+\frac{2-\a}{\a} \Delta E \left|\frac{E_{2S}}{E_*}\right|^{1-\a}.
\ee
Since the multifractal correction is going to be small, it is safe to assume that $\Delta p_*(E)\simeq\Delta E$. Then, the second term in \Eq{pfin} cannot be larger than the experimental error $\delta E$, which establishes a lower bound for the energy $E_*$:
\be\label{pfin2}
%\frac{2-\a_0}{\a_0} \Delta E \left|\frac{E_{2S}}{E_*}\right|^{1-\a_0} < \delta E\qquad \Rightarrow\qquad 
E_* >\left(\frac{\a}{2-\a}\frac{\delta E}{\Delta E}\right)^{\frac{1}{\a-1}}|E_{2S}|\,.
\ee
The smaller the experimental uncertainty $\delta E/\Delta E$ and the energies $|E_{1,2}|$ involved, the larger the lower bound on $E_*$. From Eq.\ \Eq{hyd}, the relative experimental uncertainty on the $2S$-$2P$ Lamb shift of hydrogen is $\delta E/\Delta E\approx 8.2\times 10^{-6}$ at the $3\s$ confidence level, the same as for deuterium (for helium, $\delta E/\Delta E\approx 5.5\times 10^{-5}$). The energy of the $2S_{1/2}$ state is $E_{2S}\approx-3.4\,\text{eV}$. Plugging these values into Eq.~\Eq{pfin2}, the right-hand side has a minimum at (again) $\a\approx 0.06$, resulting in the absolute lower bound $E_*> E_{\rm min}=35\,\text{MeV}$. Consistently, $|E_{2S}|/E_*\ll 1$. For the preferred value $\a=1/2$, the lower bound is much larger, $E_*>E_{\rm min}^{(\a=1/2)}=450\,\text{GeV}$. Interestingly, this is not far from the energies currently probed in the LHC.

So far, we have treated the fundamental length, time, and energy scales $\ell_*$, $t_*$, and $E_*$ in the binomial measure as independent, and $\ell_*$ has not even appeared in the analysis. A dramatic simplification of the theory takes place when all these scales are related to one another by a unit conversion. Here, the most fundamental scales of the system are those appearing in the full measures (in position and momentum space) with logarithmic oscillations, $\ell_\infty$, $t_\infty$, and $E_\infty$. Identifying these scales with the Planck scales $t_\Pl\approx 5.3912 \times 10^{-44}\,\text{s}$, $\ell_\Pl \approx 1.6163 \times 10^{-35}\,\text{m}$, and $m_\Pl \approx 1.2209\times 10^{19}\,\text{GeV}$ in four topological dimensions, we postulate that $E_*=t_\Pl m_\Pl/t_*$ and $t_*=t_\Pl \ell_*/\ell_\Pl$. Then, from the bounds $t_*<t_{\rm max}$ and $E_*>E_{\rm min}$ we have obtained on $t_*$ and $E_*$, we extract new bounds summarized in Table \ref{tab1}. For each part of the table (absolute bounds and bounds with $\a=1/2$), the ``muon lifetime'' row is $(t_{\rm max},\,\ell_{\max}:= t_{\rm max} \ell_\Pl/t_\Pl,\,\bar E_{\rm min}:= m_\Pl t_\Pl/t_{\rm max})$ while the ``Lamb shift'' row is $(\bar t_{\rm max}:=t_\Pl m_\Pl/E_{\rm min},\,\bar\ell_{\max}:= \ell_\Pl m_\Pl/E_{\rm min},\, E_{\rm min})$. In general, the Lamb-shift bounds are much stronger than those from the muon lifetime. For $\a=1/2$, the length and time scales cannot be larger than about $10^{16}-10^{17}$ Planck scales.

\begin{table}[ht]
\begin{center}
\begin{tabular}{|l|c|c|c|}\hline
Absolute bounds & $t_*$ (s)         & $\ell_*$ (m) & $E_*$ (eV) \\\hline
muon lifetime   & ${\bf <10^{-13}}$ & $<10^{-5}$   & $> 10^{-3}$ \\
Lamb shift      & $<10^{-23}$       & $<10^{-15}$  & ${\bf >10^7}$ \\\hline\hline
$\a=1/2$      & $t_*$ (s)         & $\ell_*$ (m) & $E_*$ (eV) \\\hline
muon lifetime   & ${\bf <10^{-18}}$ & $<10^{-9}$   & $> 10^{2}$ \\
Lamb shift      & $<10^{-27}$       & $<10^{-19}$  & ${\bf >10^{11}}$ \\\hline
\end{tabular}
\caption{Absolute and preferred ($\a=1/2$) bounds on the hierarchy of multifractal spacetimes with $q$-derivatives. Bounds obtained only from experiments, without the theoretical input of unit conversions, are in boldface. All figures are rounded.\label{tab1}}
\end{center}
\end{table}

These are the first stringent constraints ever obtained on the scales of the multifractal theory with $q$-derivatives. Compared with the only other extant limit $t_*< 10^6{\rm s}$ from the variation of the fine structure constant at cosmological scales \cite{frc8} (in a different multiscale theory, with weighted derivatives, but which has a very similar scale hierarchy \cite{frc2,frc13}), we have improved the experimental bounds on $t_*$ by up to 33 orders of magnitude and found brand new constraints on $\ell_*$ and $E_*$. Since $t_*$ and $\ell_*$ are much smaller than the characteristic scales of the electromagnetic ($t_{\rm QED}\sim 10^{-21}\!\!-\!\!10^{-16}\,{\rm s}$, $\ell_{\rm QED}=\infty$), weak ($t_{\rm weak}\sim 10^{-10}\!\!-\!\!10^{-6}\,{\rm s}$, $\ell_{\rm weak}\sim 10^{-18}\,{\rm m}$), and strong interactions ($t_{\rm QCD}\sim 10^{-23}\,{\rm s}$, $\ell_{\rm QCD}\sim 10^{-15}\,{\rm m}$), it is reasonable to conclude that processes involving only such forces cannot feel multiscale effects. It will be interesting to explore other physical settings, in particular cosmology, and see what experiments can further say about multifractal geometry.

%%%%%%%%%%%%%%%%%%%%%%%%%%%%%%%%%%%%%%%%%%%%%%%%%%%%%%%%%%%%%%%%%%%%%%%

\medskip

The work of G.C.\ is under a Ram\'on y Cajal contract. D.R.F.\ is supported by a GRUPIN 14-108 research grant from Principado de Asturias.

%%%%%%%%%%%%%%%%%%%%%%%%%%%%%%%%%%%%%%%%%%%%%%%%%%%%%%%%%%%%%%%%%%%%%%%
%%%%%%%%%%%%%%%%%%%%%%%%%%%%%%%%%%%%%%%%%%%%%%%%%%%%%%%%%%%%%%%%%%%%%%%

\end{document}